\documentclass[11pt]{article}
\usepackage{hyperref}
\pdfoutput=1
\begin{document}
\title{Elastic swimmer on a free surface \\ Gallery of Fluid Motion 2013\\Fluid dynamics video entry \# 102325}
\author{S. Ramananarivo, B. Thiria, R. Godoy-Diana\\
\\\vspace{6pt} Physique et M\'ecanique des Milieux Het\'erog\`enes (PMMH)\\
CNRS UMR 7636, ESPCI ParisTech, UPMC, Univ. Paris Diderot \\
10 rue Vauquelin, 75231 Paris, Cedex 5, France}
\maketitle
%% The abstract (in this file, and that submitted as text to arXiv) should include the exact phrase
%% "fluid dynamics video" or "fluid dynamics videos"
\begin{abstract}
We present in this fluid dynamics video a novel experimental setup with self-propelled swimmers on a free surface. The swimmers, modeled as flexible thin filaments, are subjected to external electromagnetic forcing driving a propagating elastic wave that gives rise to self-propulsion. The fluid-structure interaction problem of these passive anguilliform swimmers is analysed in \cite{ramananarivo2013}\end{abstract}
% main text
%\section{Introduction}
%The {\em hyperref} package is used to make links to the videos.
%%% The format is: \href{URL of video}{name that will appear in the text}
%Two sample videos are
%\href{http://ecommons.library.cornell.edu/bitstream/1813/8237/2/LIFTED_H2_EMS
%T_FUEL.mpg}{Video
%1} and
%\href{http://ecommons.library.cornell.edu/bitstream/1813/8237/4/LIFTED_H2_IEM
%_FUEL.mpg}{Video
%2}.
%It is recommended that the article include:
%\begin{enumerate}
%\item An explanation of what is shown in the video.
%\item The relevant conditions, parameters, etc..
%\item References to any papers containing further information on the
%videos.
%\item In the Abstract (in the LaTeX file and in the text submitted
%to arXiv), the exact phrase ``fluid dynamics video" or ``fluid
%dynamics videos". This is to facilitate subsequent searching.
%\end{enumerate}
%\bibliographystyle{plain}
%\bibliography{undulatory_swimming}

\begin{thebibliography}{1}

\bibitem{ramananarivo2013}
S.~Ramananarivo, R.~{Godoy-Diana}, and B.~Thiria.
\newblock Passive elastic mechanism to mimic fish-muscles action in
  anguilliform swimming.
\newblock {\em J. R. Soc. Interface}, \textbf{10}, 20130667 (2013).

\end{thebibliography}

\end{document}